\documentstyle[prd,aps,preprint,floats,epsf,rotating]{revtex}
\tightenlines
\input{epsf.tex}
\input{pstricks.tex}
\begin{document}
\hfill\parbox{4cm}{
  \hbox{CALT-68-2153             \hfill}
%  \hbox{CLNS XXXX                \hfill}
  \hbox{hep-ph/9712531              \hfill}
  \hbox{December 31, 1997           \hfill}
}

\vspace{1.5cm}
\begin{Large}
\centerline{\textrm{\bf  
Measuring Arg($V_{ub}$) in $B\to K\pi$}}
\end{Large}
\vspace{1.0cm}
\centerline{\it Frank W\"urthwein}
\centerline{\it California Institute of Technology}
\vspace{0.3cm}
\centerline{\it and}
\vspace{0.3cm}
\centerline{\it Peter Gaidarev}
\centerline{\it Cornell University}
%\vspace{1.5cm}
\vspace{0.7cm}

%\draft % makes pacs numbers print
%\large
%\title{
%\author{Frank W\"{u}rthwein}
%\date{\today}
%\maketitle

\begin{abstract}
It has previously been shown that a measurement of Arg$(V_{ub})=\gamma$
can be obtained from a triangle construction using
${\cal B}(B^0\to K^+\pi^-)$, ${\cal B}(\bar{B^0}\to K^-\pi^+)$, 
${\cal B}(B^\pm\to K^0_s\pi^\pm)$, and ${\cal B}(B^\pm\to \pi^\pm\pi^0)$.
We evaluate experimental as well as theoretical uncertainties
in this measurement. We find that a measurement of $\gamma $ with the
next generation B-factories (BaBar, BELLE, CLEO III) is
statistics rather than theory limited if
soft rescattering effects are small. 
\end{abstract}

\newpage

\section{CP violation within the Standard Model}
An ambitious experimental program to study rare decays of $B$ mesons is under 
way worldwide. The main goal of this program is to probe the Standard Model
description of quark-mixing, including CP violation. 

Within the Standard Model quark-mixing is described by a $3\times 3$
unitary matrix, the Cabibbo-Kobayashi-Maskawa (CKM) matrix.~\cite{ckm}
A unitary $3\times 3$ matrix is described by three angles (analogous to 
Euler angles) and six phases. In the standard model five of the six phases
are non-physical. They can be eliminated by redefining the phases of the
quark fields. 

The one remaining phase could in principle provide for a very rich set
of CP violating phenomena.
However, nature, as we know it, appears to have chosen the angles such that 
all but two of the matrix elements may
have approximately the same phase,
zero by convention.
The two matrix elements that stand out
are $V_{td}$ and $V_{ub}$.

At present, the experimentally preferred ranges~\cite{buras} for the phases
of these elements are given by:
\begin{equation}\begin{array}{rcccl}
41^\circ &\le & \mathrm{Arg}(V_{ub}) &\le & 134^\circ \\
11^\circ &\le & \mathrm{Arg}(V_{td}) &\le & 27^\circ\\
\end{array}\end{equation} 

They are obtained using the unitarity constraint
$V_{ud}V^*_{ub}+V_{cd}V^*_{cb}+V_{td}V^*_{tb}=0$, in combination
with measurements of $|V_{ud}|,\ |V_{cd}|$, as well as
$B^0-\bar{B^0}$ mixing, $|V_{cb}|$ and $|V_{ub}|$ from
semi-leptonic $B$ decays, and CP violation in $K_L\to\pi\pi$.

The factor of six
difference in the preferred region for the two phases is
mostly due to the geometry of this ``unitarity triangle'' rather than
to the difference in errors on 
$|V_{td}|=(8.6\pm 1.1)\times 10^{3}$~\cite{buras}
and $|V_{ub}|= (3.2\pm 0.8)\times 10^{-3}$~\cite{pdg96}.
%$6.9<|V_{td}|\times 10^{3}< 11.3$~\cite{buras}

A number of experiments
expect to measure $\sin 2\beta$\ ($\beta = $Arg$(V_{td})$)
within the next few years. 
Each one of them expects to reach a sensitivity somewhat better than
the precision we currently have via indirect 
means.
%as can be seen in Table~\ref{tab:beta}~\cite{beta}.
While this will already provide for a stringent test of the standard model, 
one would ideally want to measure Arg$(V_{ub})$ to similar precision
to verify that the phase structure of nature's preferred quark-mixing
matrix is consistent with coming from a single phase.

There is certainly no dearth of techniques for measuring 
Arg$(V_{ub})$~\cite{measuringVub}. Unfortunately, most techniques
either require $B_s$ decays or rather large number of $B_{d,u}$ decays.
One of the experimentally 
more promising techniques to measure Arg$(V_{ub})=\gamma$
was proposed by Fleischer~\cite{fltriangle}.
In his paper, Fleischer acknowledges that his method can only
provide a rough estimate of $\gamma$ without attempting to
quantify the precision obtainable.
We show in the present paper that this theoretical 
error is small compared to
the experimental error (at least in the near future)
for a large part of the allowed parameter space.

This work is organized as follows. Section II gives an overview of the method,
as well as a justification for the range of parameters used throughout
the paper. Section III discusses the theoretical uncertainty on
Arg($V_{ub}$) using this method.
Section IV provides estimates of the expected experimental error for
one ``nominal year'' of PEP-II and CESR phase 4.
%Sections V discusses the expected sensitivity for measuring $\gamma$ if
%strong phases are small. 
And Section V discusses the experimental
sensitivity towards measuring direct CP violation in $B\to K^\pm\pi^\mp$.
In particular, we show that for CESR phase 4 luminosities we will
either measure direct CP violation or Arg($V_{ub} $) to within reasonable
theoretical uncertainties.
Summary and conclusions are given in Section VI.

After completion of this work we have learned of a paper by 
Gronau and Rosner~\cite{grnew}
that also discusses this triangle from a somewhat different perspective.

%\begin{table}
%\caption{Expected sensitivity 
%towards measuring Arg$(V_{td})=\beta $.[4] }
%\label{tab:beta}
%\begin{tabular}{lcc}
%Experiment & $\delta\sin 2\beta$ & years of nom. lum. \\
%\hline
%BaBar & 0.11-0.14 & 1 \\
%BELLE & 0.12 & 1 \\
%CDF   & 0.08-0.13 & 2 \\
%D0    & 0.12 & 1 \\
%Hera-B & 0.13 & 1 \\
%Hera-B & 0.065 & 4 \\
%BTeV   & 0.04  & 1 \\
%\end{tabular}
%\end{table}

\section{Measuring $\gamma$ in $B\to K\pi$}
The decay amplitudes for the decays $B_d\to K^\pm\pi^\mp$ and
$B_u\to K^0\pi^\pm$ are given by:
\begin{equation}
\begin{array}{ccc}
A_{B^+\to K^0\pi^+} & = & V_{ts} V^\star_{tb}P_{ts} + V_{cs}V^\star_{cb} 
P_{cs}  \\
A_{B^0\to K^+\pi^-} & = & -(V^\star_{ub} V_{us} T_{s} + 
V_{ts} V^\star_{tb}P_{ts} + V_{cs}V^\star_{cb} P_{cs}) 
\\%& =: T_{K} e^{i(\Delta\phi +\gamma)} + P_{s} \\
A_{\bar{B^0}\to K^-\pi^+} & = & -(V_{ub} V^\star_{us} T_{s} + 
V^\star_{ts} V_{tb}P_{ts} + V^\star_{cs}V_{cb} P_{cs}) 
\\ %& =: T_{K} e^{i(\Delta\phi -\gamma)} + P_{s} \\
\label{eq:ampkp}\\
\end{array}
\end{equation}

Here $T_s,P_{cs,ts}$ refer to the $\Delta S = 1$
external W-emission and gluonic penguin ($c,t-$quark in the loop) amplitudes
respectively. 
There are in principle also contributions from color-suppressed electroweak
penguin, and annihilation diagrams. These 
are expected to amount to at most a few percent, and
are therefore ignored here.\cite{fkwcbx}

Recently, it has been
pointed out~\cite{fsiphases} that
soft final-state interactions may drastically enhance 
contributions from annihilation diagrams. 
Furthermore, arguments based on Isospin~\cite{fsiphases} imply that these 
contributions can in general not be neglected.
The large samples of
$\Upsilon (4S)$ decays expected at BaBar, BELLE, and CLEO in the next few 
years will allow us to decide this question 
experimentally~\cite{grnew,fsikagan,fsifkw}.
%~\cite{grnew}~\cite{fsifkw}.

\begin{figure}
\centering
\leavevmode
\epsfxsize=15cm
\epsfbox{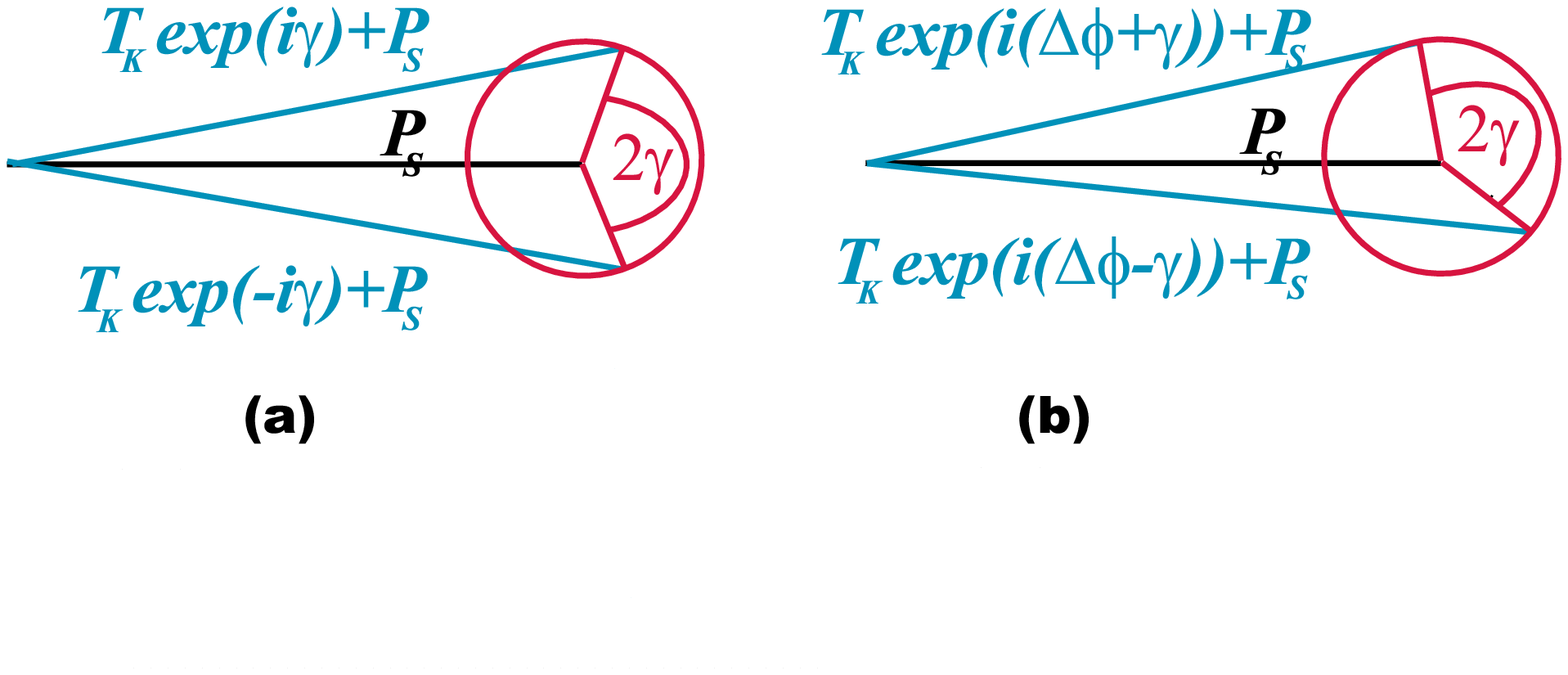}
\vspace{0.4cm}
\caption{Triangle construction for measuring Arg$(V_{ub})=\gamma$ as
described in the text. $T_K$ determines the radius of the circle.
The triangles are drawn to scale, assuming $T_K/P_S = 0.2$.}
\label{fig:fltri1}
\end{figure}

Neglecting soft final-state interactions,
the only non-trivial weak phase in these decays 
is Arg$(V_{ub}) = \gamma$.
The triangle constructions in Figure~\ref{fig:fltri1} thus allow for the
determination of Arg($V_{ub}$) up to a fourfold ambiguity from the
following time
integrated decay rate measurements if 
$\sqrt{\tau_B}\times |V_{ub} V^\star_{us} T_{s}| = T_K$ is known.
\begin{equation}
\begin{array}{ccc}
\sqrt{{\cal B}(B^-\to \bar K^0\pi^- )} = \sqrt{{\cal B}(B^+\to K^0\pi^+ )}
= \sqrt{\tau_B}\times |A_{B^+\to K^0\pi^+}| &=& P_S \\
\sqrt{{\cal B}(B^0\to K^+\pi^- )} &=& 
|T_K e^{i(\Delta\phi -\gamma)} + P_{S}|  \\
\sqrt{{\cal B}(\bar B^0\to K^-\pi^+ )} &=& 
|T_K e^{i(\Delta\phi +\gamma)} + P_{S}|  \\
\end{array}
\end{equation}

Here $\Delta\phi$\ is the strong phase difference between the sum of the two
penguin amplitudes and the 
external W-emission amplitude. 
Figure~\ref{fig:fltri1}(a) illustrates the
special case $\Delta\phi = 0$. The circle depicts $T_K$.
Allowing for $\Delta\phi\neq 0$ results in a rotation as shown
in Figure~\ref{fig:fltri1}(b).
This rotation produces direct
CP violation, meaning a rate asymmetry between $B^0\to K^+\pi^-$
and $\bar{B^0}\to K^-\pi^+$.

The size of $T_K$ may be determined either from $B\to\pi l\nu$, assuming
factorization, or from $B^\pm\to\pi^\pm\pi^0$, correcting for the extra
contribution due to color-suppressed W-emission in this decay. 
Both of these determinations are
plagued by theoretical uncertainties at the $10-20\%$ level.

\newpage

Assuming factorization, we can relate $T_K$ to $B\to\pi l\nu$ at
$q^2=m_{\pi^2}$ via:~\cite{lkgpilnu}
\begin{equation}
\begin{array}{ccccccccc}
T_K &
\sim &
\sqrt{6}\pi f_K |V_{us}| &
\times & 
a_1 & 
\times & 
\sqrt{d\Gamma(B\to\pi l\nu)/dq^2 |_{q^2 = m_{\pi}^2}
\over \Gamma(B\to\pi l\nu)} &
\times &
\sqrt{{\cal B}(B\to\pi l\nu)} \\
 &\sim & 0.27\mathrm{GeV} &\times &(1.0\pm 0.1) &\times & 
(0.27\pm 0.05)/\mathrm{GeV} &\times & 
(0.0135\pm 0.0022)\\
 &=& (9.8\pm 2.5)\times 10^{-4}
\end{array}
\label{eq:tkpilnu}
\end{equation}

Our present estimate of $T_K$ is limited by the
statistics of the experimental data in $B\to\pi l\nu$.
However, in a few years this error due to statistics
will be small compared to the error
on $a_1$. 
For $b\to c$ transitions, theory~\cite{burasa1} ($a_1 = 1.01\pm 0.02$)
and experiment~\cite{jorge} ($a_1 = 1.03\pm 0.07$) agree quite well.
However, it is far from obvious to us how well this will extrapolate
from ``heavy-to-heavy'' ($b\to c$)
to ``heavy-to-light'' ($b\to u$) transitions. 
We therefore set $a_1 = 1.0$ with a 10\% error.
We expect this to be the dominant source of uncertainty in the future.

Unfortunately,
the situation in
$b\to u$ decays is different from $b\to c$ decays in that
there are no non-leptonic decays for which only contributions from 
external W-emission are relevant. The closest one can come
to an experimental measurement
of $T_K$ in non-leptonic $B$ decays is via:
\begin{equation}
T_K \sim \sqrt{2{\cal B}(B^\pm\to \pi^0\pi^\pm )}\times f_K/f_\pi
\times |V_{us}/V_{ud}|\times a_1/(a_1+a_2) 
\label{eq:tkpipi}
\end{equation}
 
The factor $a_1/(a_1+a_2)\approx 1/1.3$ 
enters because $B^\pm\to\pi^0\pi^\pm$ has 
contributions from color-allowed as well as color-suppressed W-emission
amplitudes, whereas $T_K$ refers to the color-allowed transition only.
Again, one can infer $a_1/(a_1+a_2)$ in $b\to u$ decays from a measurement
in $b\to c$ decays.

So far, CLEO has measured only
${\cal B}(B^+\to K^+\pi^0 + B^+\to\pi^+\pi^0 ) = 
(1.6^{+0.6}_{-0.5}\pm 0.3\pm 0.2)\times 10^{-5} $.~\cite{kpiprl} 
However, within
the next few years CLEO, BaBar, and BELLE will each be able to measure
$\sqrt{{\cal B}(B^\pm\to\pi^\pm\pi^0)}$ to within $5-10\%$.
While $a_1/(a_1+a_2)$ may eventually be measured quite precisely
in $b\to c$ decays, it is again the extrapolation from $b\to c$ to $b\to u$
that is likely to be the dominant source of uncertainty
in determining $T_K$ in this fashion.

In principle, uncertainties due to
non-factorizable SU(3) symmetry breaking effects also affect this
method for estimating $T_K$. However, they are probably small
compared to the uncertainties in $a_1$ and $a_1/(a_1+a_2)$ respectively.

Using the recent CLEO result~\cite{kpiprl} 
${\cal B}(B^+\to K^0\pi^+) = (2.3^{+1.1}_{-1.0}\pm 0.3\pm 0.2)\times 10^{-5}$
for $P_S^2$, and Equation~(\ref{eq:tkpilnu})
our present best guess for
$T_K/P_S$ is given by:

\begin{equation}
T_K/P_S \sim {(9.8\pm 2.5)\times 10^{-4}\over (4.8\pm 1.2)\times 10^{-3}}
= 0.20\pm 0.07 
\end{equation}

In the future, 
combining Equations~(\ref{eq:tkpilnu}) and (\ref{eq:tkpipi}) 
may eventually allow for a determination of
$T_K$ to within $\pm (10-20)\%$.

\section{Theoretical error on $\gamma$.}

In the following, we present estimates of the error on the measured
Arg($V_{ub}$) as a function of the geometry of the 
$K\pi $ amplitude triangles, the true
value for $\gamma$, the relative error on $T_K$ 
($\delta T_K/T_K = 
(T_K(\mathrm{theory})-T_K(\mathrm{nature}))/T_K(\mathrm{nature})$),
and the strong phase difference $\Delta\phi$. 
We choose $T_K/P_S = 0.2$ throughout. However, we verified that varying
$T_K/P_S $ within a reasonable range has only a small effect.

The point of this exercise is to ``propagate'' the theoretical error
on $T_K$ into an error on Arg($V_{ub}$), and show how this error
propagation depends on nature's choice for the size of
the relevant quantities involved.

\subsection{Geometry of the Triangles}

\begin{figure}
\centering
\leavevmode
\epsfxsize=12cm
\epsfbox{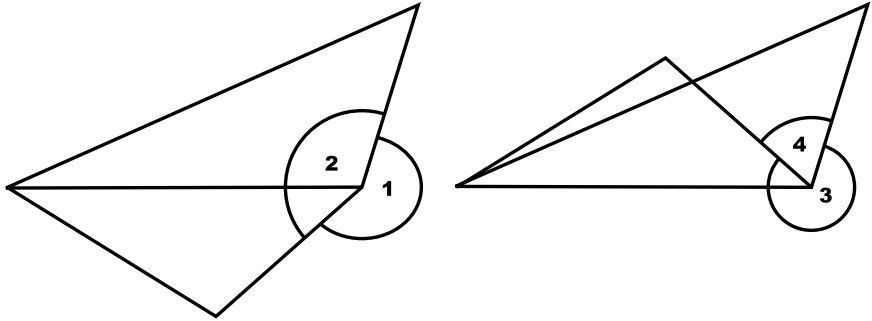}
\caption{Depiction of fourfold ambiguity.}
\label{fig:fltri2}
\end{figure}

Let us first look at the fourfold ambiguity in determining $\gamma$
via the $ K\pi $ amplitude triangles.
A strong phase difference $\Delta\phi = 180^\circ$ produces the same
triangles as $\Delta\phi = 0$, except that now $2\gamma$ is given by the 
angle inside the two triangles. This ambiguity of interpreting the inside or 
outside angle as $2\gamma$ makes up half of the fourfold ambiguity
in extracting $\gamma$ in this fashion. The other half is given by
flipping one of the two triangles around the baseline as shown in 
Figure~\ref{fig:fltri2}.

\begin{figure}
\centering
\leavevmode
\epsfxsize=15cm
\epsfbox{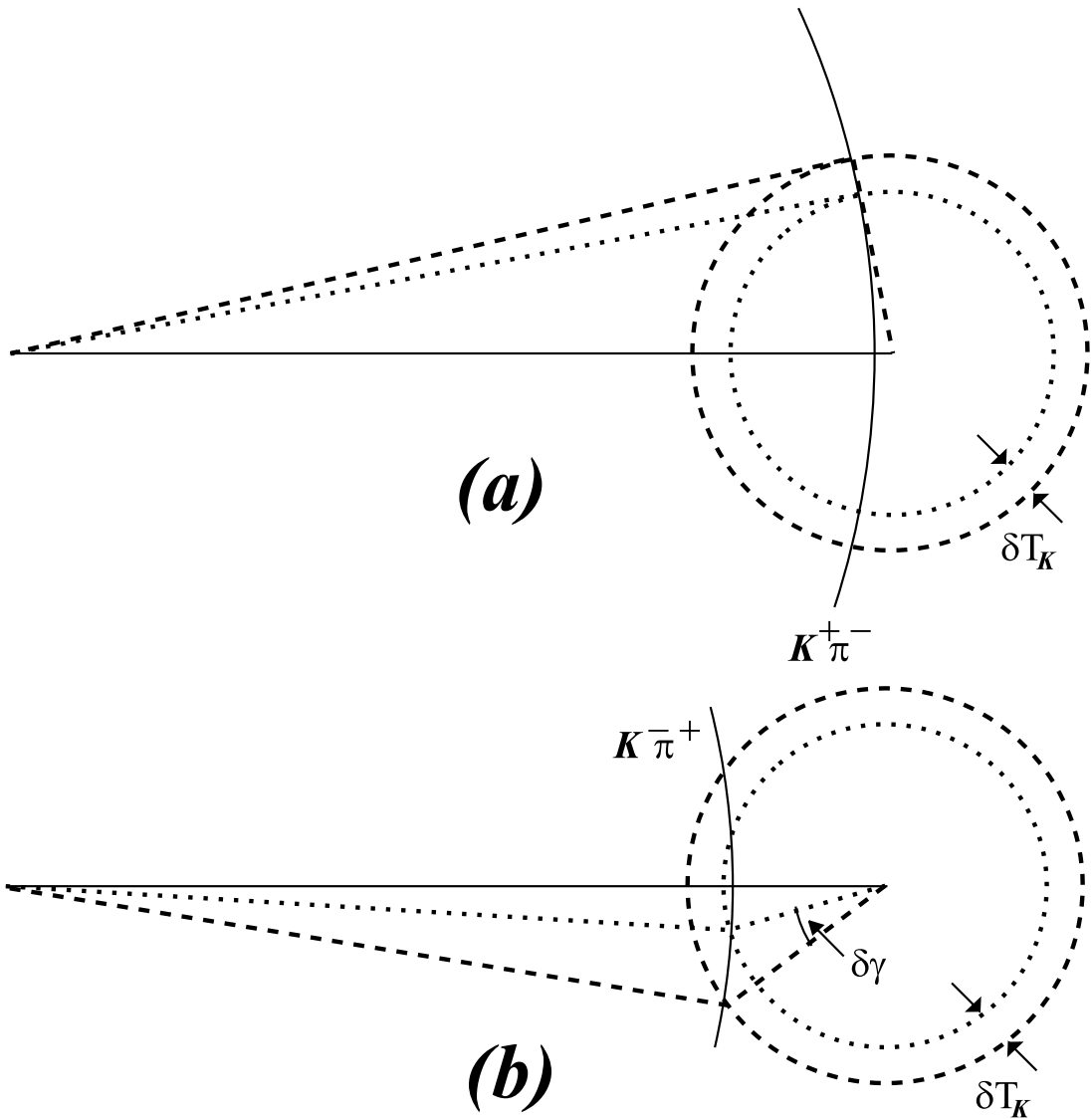}
\vspace{0.4cm}
\caption{Geometry for which $\delta T_K\ne 0$ minimally ((a))
and maximally ((b)) affects the measurement of $\gamma$.
The dashed and dotted lines correspond to a change in $T_K$ by $20\%$.
The resulting change ($\delta\gamma$) in the measured $\gamma$ 
is negligible for the geometry illustrated in Figure (a).} 
\label{fig:fltri3}
\end{figure}

The effect of $\delta T_K$
on the measured $\gamma$ strongly depends on the geometry of
the two triangles. 
This is shown in
Figure~\ref{fig:fltri3}(a) and (b). 
An error on $T_K$ will affect $\gamma$
most if one or both of the two triangles
are close to degenerate (\ref{fig:fltri3}(b)), and least if
$T_K$ is close to tangential to one or both of the circles with radius 
$\sqrt{{\cal B}(B^0\to K^+\pi^- )}$ and 
$\sqrt{{\cal B}(\bar B^0\to K^-\pi^+ )}$ (\ref{fig:fltri3}(a)). 
For $\delta T/T < 0.0$ 
one can easily imagine a situation where one or both 
triangles no longer close.

\begin{figure}
%\centering
\leavevmode
\epsfxsize=7.5cm
\epsfbox[45 150 525 630]{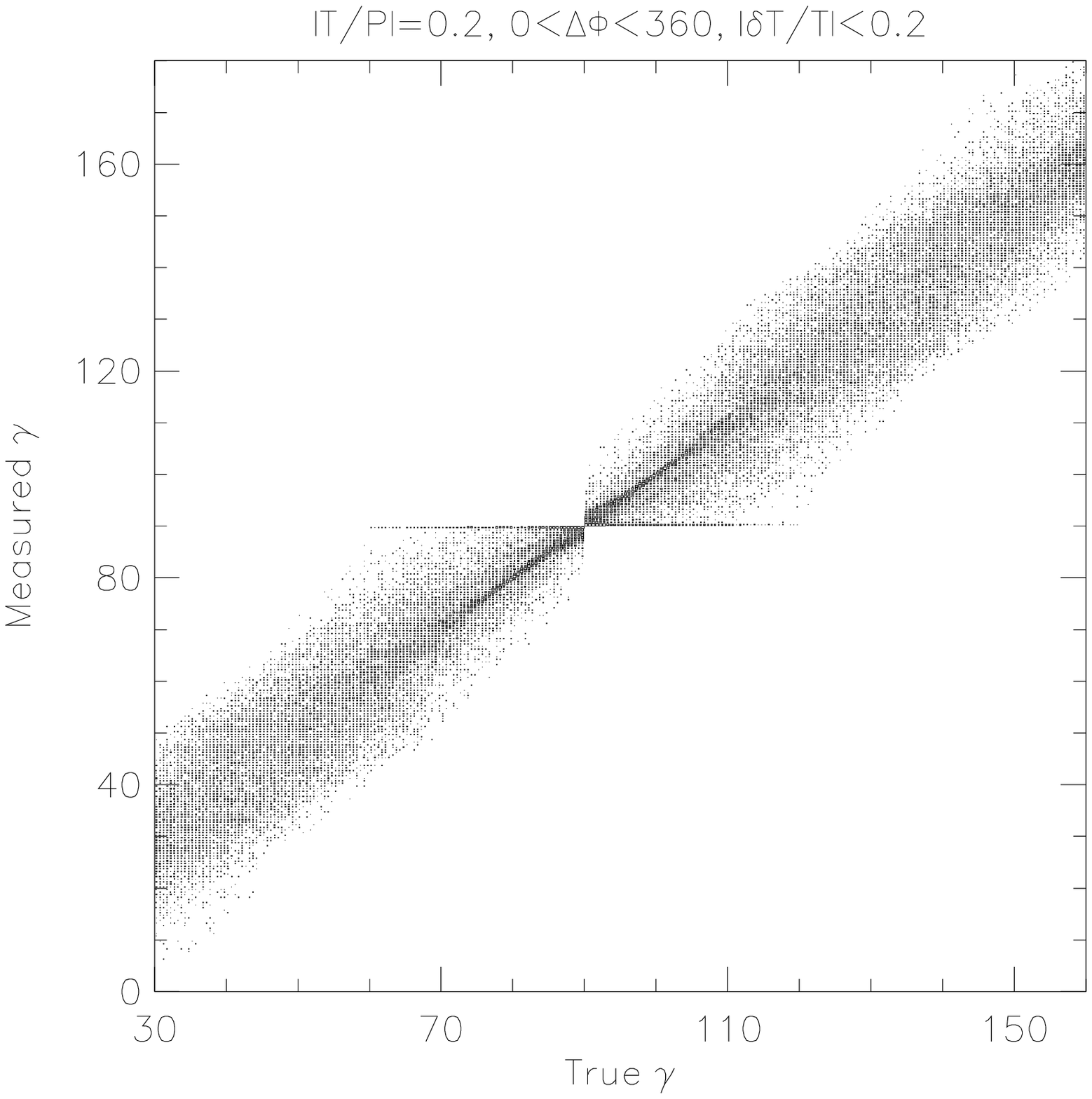}
\hspace{0.5cm}
%\caption{Measured versus true $\gamma$ for 
%$T_K/P_S =0.2$,
%and $|\delta T_K/T_K|<\pm 0.2$.}
%\label{fig:dphinorestrict2}
%\end{figure}
%\begin{figure}
%\centering
%\leavevmode
\epsfxsize=7.5cm
\epsfbox[45 150 525 630]{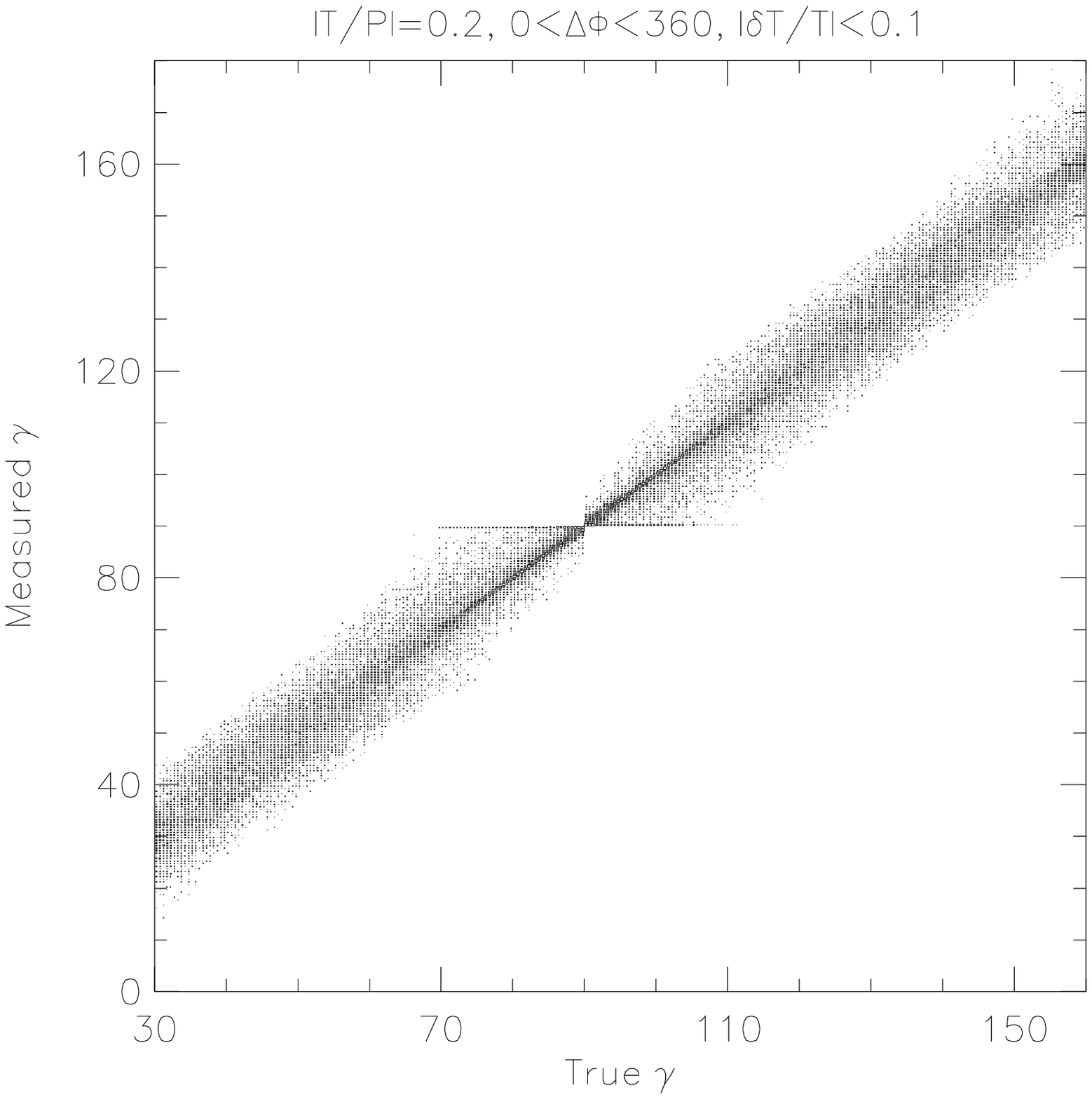}
\vspace{0.2cm}
\caption{Measured versus true $\gamma$ for 
$T_K/P_S =0.2$, and $|\delta T_K/T_K|<\pm 0.2$ (left) or  
$|\delta T_K/T_K|<\pm 0.1$ (right).}
\label{fig:dphinorestrict}
\end{figure}

%%%%%%%%%%%%%%%%%%%%%%%%%%%%%%%%%%%
%Figure~\ref{fig:dphinorestrict2} and 
Figure~\ref{fig:dphinorestrict} 
shows 
measured versus true $\gamma$ for $|\delta T_K/T_K|<0.2$ (left)
and $|\delta T_K/T_K|<0.1$ (right) respectively. 
No restrictions on the geometry are
applied here. 

These figures are generated as follows. We assume infinite statistics
for the branching fraction measurements. We then choose some ``true''
value for
$\gamma$, $\delta T_K/T_K$, and $\Delta\phi$.
These are drawn randomly from flat distributions within the specified
ranges ($30^\circ \le \gamma\le 160^\circ $, 
$0^\circ\le \Delta\phi\le 360^\circ$, $|\delta T_K/T_K |\le 0.1,0.2$).
Having fixed these parameters we then ``measure'' $\gamma $.
In the end,
each point in these figures corresponds to a possible pair of measured and
true $\gamma$.

We show only one of the four possible solutions for 
$\gamma$ in these figures. 
%resolve the fourfold ambiguity by always
%choosing the option closest to the true $\gamma$. 
A second set of solutions
is given by the reflection of the figure along the axis 
true $\gamma = 90^\circ$. 
The third and fourth solution are not very 
illuminating as they correspond to measuring $\Delta\phi$ instead of $\gamma$
within a twofold ambiguity. 

The ``line'' at measured $\gamma = 90^\circ$ is due to the fact
that we set $\gamma = 90^\circ$ if neither of the triangles closes.

\begin{figure}
\centering
\leavevmode
\epsfxsize=7.5cm
\epsfbox[45 150 525 630]{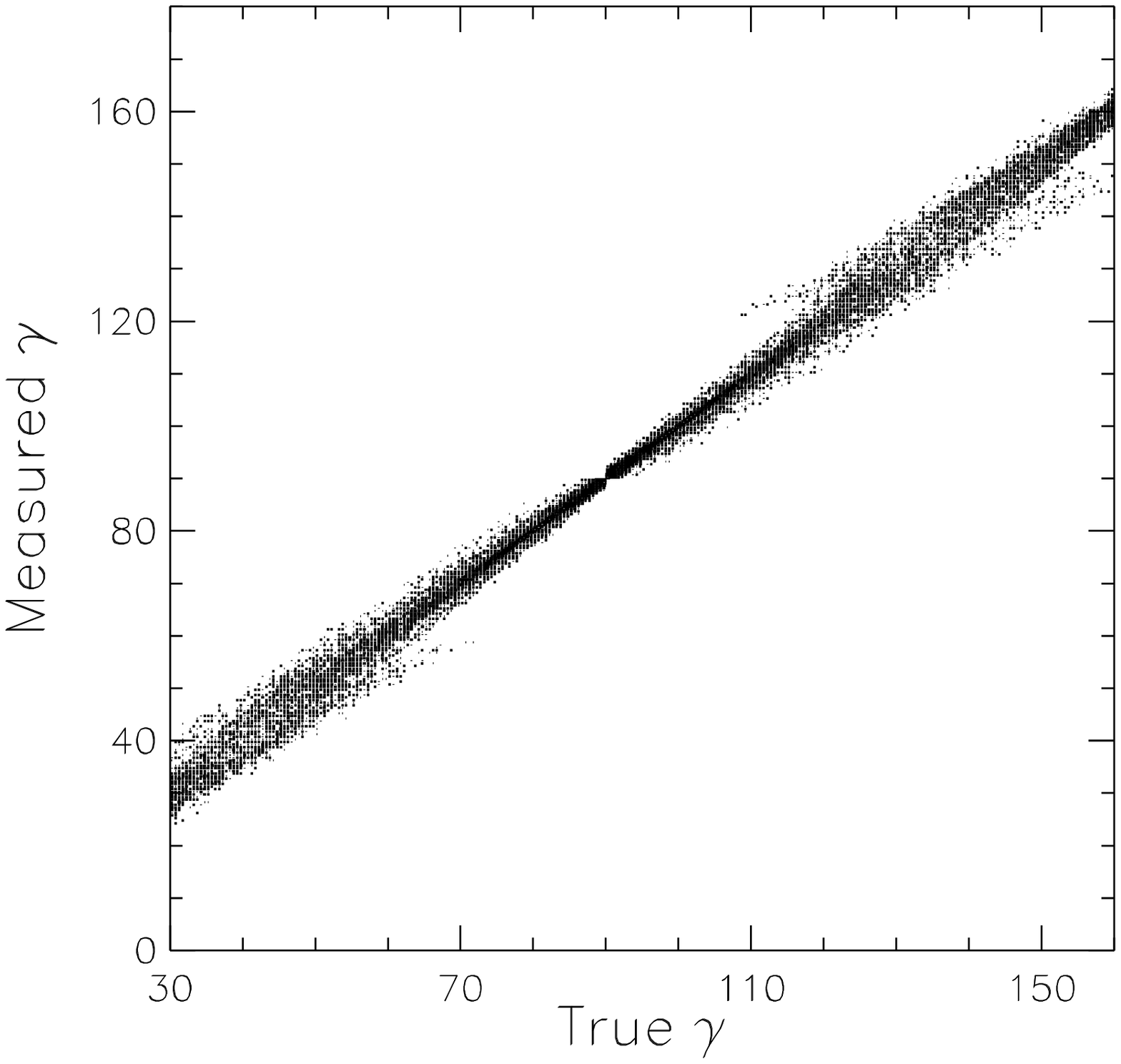}
\caption{Measured versus true $\gamma$ for 
$T_K/P_S =0.2$, $|\delta T_K/T_K|<\pm 0.1$, and $|\sin\theta_i|>0.5$.}
\label{fig:restrictgeo}
\end{figure}

There appears to be a $20-30$ degree range of values for
the true $\gamma$ for any given measured $\gamma$.
However, this is somewhat misleading as can be seen in 
Figure~\ref{fig:restrictgeo}. Here we assumed that nature is kind to us
experimentalists in that 
the measured angles ($\theta_i,\ i=1,2$) between $T_K$ and $P_S$ for the two 
triangles turn out to be in the range $|\sin\theta_i|>0.5$, and
$|\delta T_K/T_K |<0.1$. In other words, nature's choice is a geometry
closer to Figure~\ref{fig:fltri3} (a) rather than (b).

The range of possible true values for a given measured
value of $\gamma$ then reduces from about $25^\circ$ to $10^\circ-15^\circ$.
In this case the four possible solutions for $\gamma$ tend to be clearly
separated. This is unfortunately not the case for 
%Figure~\ref{fig:dphinorestrict2} and 
Figure~\ref{fig:dphinorestrict}.

%In the following, we show some reasonable predictions for the expected
%experimental error. After that we discuss
%two possible physics scenarios that
%would naturally provide a restriction of the allowed geometry, given
%the experimentally preferred range of $41^\circ < \gamma < 134^\circ$. 

\subsection{Assuming $\Delta\phi$ is small}

It has often been stated
that the large $q^2$ in $B$ decays to $K\pi$
leads to rather small final state interactions, and thereby to 
very small $\Delta\phi$. The case of small $\Delta\phi$ is therefore
of special interest.

\begin{figure}
\centering
\leavevmode
\epsfysize=7cm
\epsfbox[45 150 525 630]{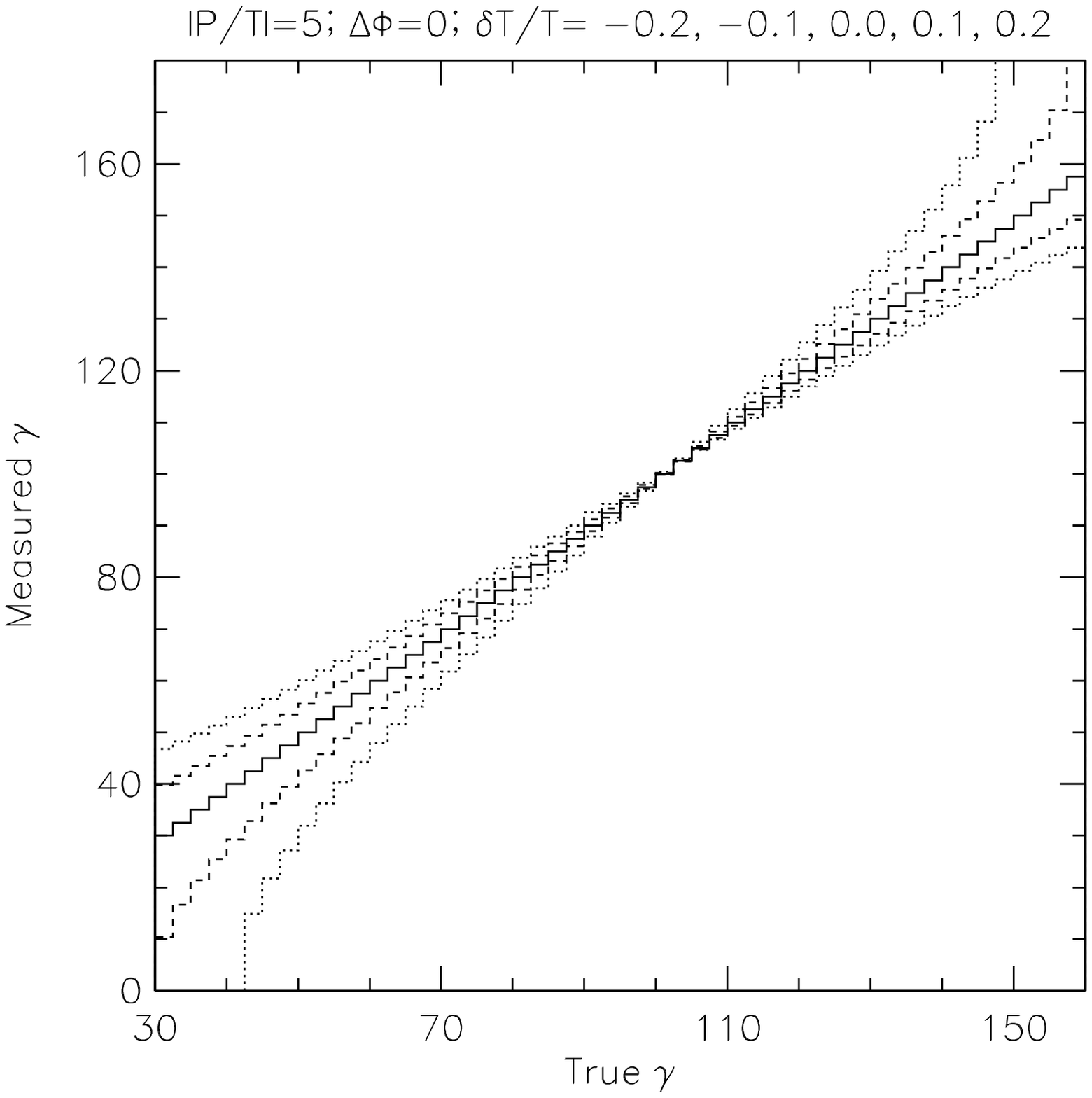}
\caption{Measured versus true $\gamma$ for $\Delta\phi = 0$, $T_K/P_S =0.2$,
and $\delta T_K/T_K = 0$(solid), $\pm 0.1$(dashed), and $\pm 0.2$(dotted)}
\label{fig:dphizero}
\end{figure}

\begin{figure}
\centering
\leavevmode
\epsfysize=7cm
\epsfbox[45 150 525 630]{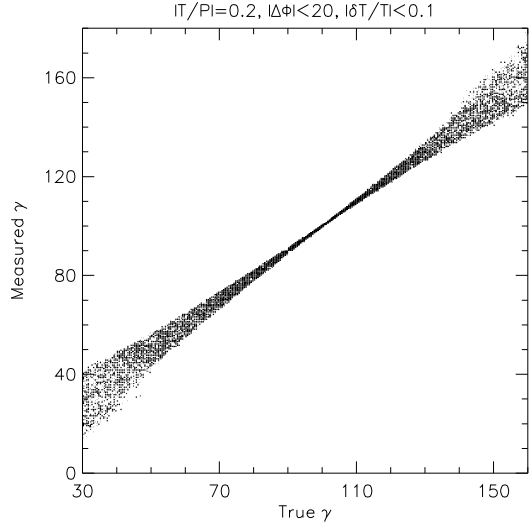}
\caption{Measured versus true $\gamma$ for 
$|\Delta\phi| < 20^\circ$, $T_K/P_S =0.2$,
and $|\delta T_K/T_K|<\pm 0.1$.}
\label{fig:dphirestricted}
\end{figure}

For the experimentally preferred range of $41^\circ < \gamma < 134^\circ$ 
we are in general
less sensitive to $\delta T/T\neq 0$ if $\Delta\phi \sim 0$.
This is immediately obvious from the discussion of the geometry above.
To get a degenerate triangle requires $\Delta\phi = \pm\gamma$ for 
$\cos\gamma >0.0$, or $\Delta\phi = \pm(180^\circ -\gamma)$ for
$\cos\gamma <0.0$. 
If strong phase differences are indeed small 
%as generally expected,
then this means that $\gamma\sim 100^\circ$ is largely free of uncertainties
due to $\delta T_K/T_K$. This is shown graphically in
Figures~\ref{fig:dphizero} and \ref{fig:dphirestricted}. 

Figure~\ref{fig:dphizero} 
shows the envelopes for $\delta T_K/T_K = \pm 0.1$ and
$\pm 0.2$ for $\Delta\phi = 0.0$. 
And Figure~\ref{fig:dphirestricted} depicts the range of possible measured
versus true $\gamma$ for $|\delta T_K/T_K|<0.1$ and $|\Delta\phi |<20^\circ$.
For a significant 
part of the currently preferred range ($ 70^\circ \le \gamma\le 130^\circ$) 
the 
theoretical error is less than about $\pm 5^\circ$ if 
$|\Delta\phi|<20^\circ$. This compares well with the expected experimental 
error presented in the next section.

\section{Error on $\gamma $ due to statistics}
CLEO has recently measured~\cite{kpiprl} 
${\cal B}(B^0\to K^+\pi^-) = (1.5^{+0.5}_{-0.4}\pm 0.1\pm 0.1)\times 10^{-5}$
and
${\cal B}(B^+\to K^0\pi^+) = (2.3^{+1.1}_{-1.0}\pm 0.3\pm 0.2)\times 10^{-5}$.
In both cases averaging over charge conjugate states is implied.
These measurements use $3.1$fb$^{-1}$ of integrated luminosity, taken
at the $\Upsilon (4S)$ resonance. 

The design luminosity for PEP-II and CESR phase 4 is $30$fb$^{-1}$ and 
$300$fb$^{-1}$ per $10^7$ seconds, with PEP-II starting operations 
in 1999, and CESR phase 4 being planned for 2003. 
An estimate of the future error on the 
relevant $B\to K\pi$ 
branching fractions may be obtained as follows.
We repeat the recent CLEO analysis for $B\to K^\pm\pi^\mp$
on Monte Carlo generated data that has
the $K\pi$ particle ID separation dialed up to $4\sigma$ but is otherwise
modelled to reflect the distributions found in CLEO data.
We do this for 1,000 Monte Carlo experiments, allowing 
for Poisson fluctuations of the number
of generated signal events in $\pi^+\pi^-$ and $K^\pm\pi^\mp$ 
around the central values as determined in data.
We measure
the error on ${\cal B}(B\to K^\pm\pi^\mp)$ for each experiment and
determine the mean error, averaged over the 1,000 experiments.
We then scale this mean
error on the branching fraction by $1/\sqrt{5}$ and $1/\sqrt{50}$
respectively to arrive at an expected
error of $10\%$ and $3\%$ for one ``nominal'' 
year of PEP-II and CESR phase 4.
The measured yield in $K^0_s\pi^\pm$ is roughly half that of $K^\pm\pi^\mp$.
We therefore 
expect ${\cal B}(B^\pm\to K^0_s\pi^\pm)$ to be measured with roughly
the same precision as either one of the charge conjugate modes 
for $B\to K^\pm\pi^\mp$. 

We can now repeat the Monte Carlo exercise from the previous section,
allowing for a Gaussian error on all sides, but no theoretical error on
$T_K$. We assume the relative error on $T_K$ to be the same as that for
the other sides. This is equivalent to assuming a yield in $B\to \pi^\pm\pi^0$
half that for $B\to K^\pm\pi^\mp$. This is 
consistent with the latest CLEO results.
The resulting 
Gaussian error on
$\gamma$ is $17^\circ$, and $8^\circ$ for
a $10\%$, and $3\%$
error on the measured branching fractions. 

%%%%%%%%%%%%%%%%%%%%%%%%
\section{Connection to direct CP violation in $B\to K^\pm\pi^\mp$}
%%%%%%%%%%%%%%%%%%%%%%%%
\label{seq:asym}

In the section on the geometry of the triangle we have shown that
the error on $\gamma$ is largest if one of the two triangles is degenerate.
Given the experimentally preferred range of $41^\circ <\gamma < 134^\circ$
such a geometry tends to imply a large CP violating asymmetry in
$B\to K^\pm\pi^\mp$. In this section we want to explore this a little further
by showing that (at least for CESR phase 4 type luminosities)
we are in a ``win-win'' situation for a large part of the
experimentally preferred range in $\gamma$. We either measure 
direct CP violation or $\gamma$ with reasonably small theoretical 
uncertainties.

In general:
\begin{equation}
A_{cp} =
{2 T_K\ P_S\times \sin\Delta\phi\sin\gamma 
\over
T_K^2 + P_S^2 + 2\ T_K P_S\times \cos\Delta\phi\cos\gamma }\\
= 2\sin\Delta\phi\sin\gamma\times {T_K\over P_S} + O((T_K/P_S)^2)
\end{equation}

For example, for $\gamma\sim 90^\circ$ and $T_K/P_S\sim 0.2$
a limit of about $A_{cp} < 15\%$ would 
constrain $\Delta\phi<20^\circ$, the value chosen 
for Figure~\ref{fig:dphirestricted}.
%in the previous section.

%Given that we expect $T_K/P_S\sim 0.2$
%a CP violating
%rate asymmetry as large as about $40\% $ is not unreasonable within the 
%Standard Model.

Let us briefly look at the expected sensitivity for measuring
direct CP violation in $K^\pm\pi^\mp$.
CLEO uses a Maximum Likelihood fit
to measure
the yield in $B^0\to K^+\pi^-$ ($N_1$) and 
$\bar{B^0}\to K^-\pi^+$ ($N_2$).
The error on the asymmetry is then
given by:
\begin{equation}
\sigma_{A_{cp}}^2 = {4\over (N_1+N_2)^4}\times 
((N_2\ \sigma_{N_1})^2 + (N_1\ \sigma_{N_2})^2 
- 2 N_1N_2\rho\ \sigma_{N_1}\sigma_{N_2} )
\label{eq:sigacp}
\end{equation}

The errors on $N_i$ ($\sigma_{N_i}$) are in general
larger than $\sqrt{N_i}$ because of backgrounds from continuum as well
as $B\to\pi^+\pi^-$, and $K/\pi$ double miss-id.
This also leads to a non-zero value of the correlation coefficient
($\rho$) between $N_1$ and $N_2$. 
In general, $K\pi$ double miss-id causes $\rho < 0$ whereas
CP symmetric backgrounds lead to $\rho > 0$.

At present $\rho \sim -0.2$. Using Monte Carlo
we can dial up the $K/\pi$ separation to $4\sigma$ to simulate a
CLEO III analysis. This results in $\rho \sim -0.008$.
To scale $\sigma_{N_i}$ we use:
\begin{equation}
\sigma_{N_i}^2 = N_i + B/S\times {(N_1 + N_2)\over 2} 
\label{eq:sigmani}
\end{equation}

For a counting analysis, $B/S$ is just the expected background to signal
ratio. In a multi-dimensional likelihood fit like the CLEO analysis,
one generally includes large sidebands in the fit. There is therefore
no simple definition of $B/S$ because there is no obvious signal box defined.
We therefore define an ``effective $B/S$'' via:
\begin{equation}
B/S = \langle {\sigma^2_{N_1+N_2} \over N_1+N_2} - 1 \rangle
\end{equation}

The average is formed over many Monte Carlo generated 
experiments of $3.1$fb$^{-1}$ integrated
luminosity each.
%, generated via Monte Carlo
%according to the distributions found in CLEO data.  
When we increase the $K/\pi$ particle ID separation in Monte Carlo
to $4\sigma$ we find 
$B/S = 0.24$. For the current
CLEO analysis ($K\pi$ separation $\sim 1.7\sigma @ p\sim 2.6$GeV$/c$)
this effective $B/S$ is about a factor of three worse.
To determine $B/S$ we assume 
${\cal B}(B\to\pi^+\pi^-)/{\cal B}(B\to K^\pm\pi^\mp) \sim 0.5$. 
However, the resulting value for $B/S$ depends only very weakly on 
this ratio.

Using Equation (\ref{eq:sigmani}) we can rewrite Equation (\ref{eq:sigacp})
into a more useful form:
\begin{equation}
\begin{array}{rccl}
\sigma_{A_{cp}}^2 &=& ( &
 (1+A_{cp})^2\times (1-A_{cp}+B/S) \\
 & & + & (1-A_{cp})^2\times (1+A_{cp}+B/S)\\ 
 & & - & 2\rho \times (1-A_{cp}^2)\times \sqrt{(1+B/S)^2 - A_{cp}^2}\ \ \ ) \\
 & & / & 2\epsilon \times {\cal L} \times {\cal B}(B\to K^\pm\pi^\mp)\\
\end{array}
\label{eq:sigacpsqr}
\end{equation}

\begin{figure}
\centering
\leavevmode
\epsfxsize=12cm
%\epsfbox[45 150 525 630]{sigacp_cleo3.ps}
\begin{turn}{-90}
\epsfbox{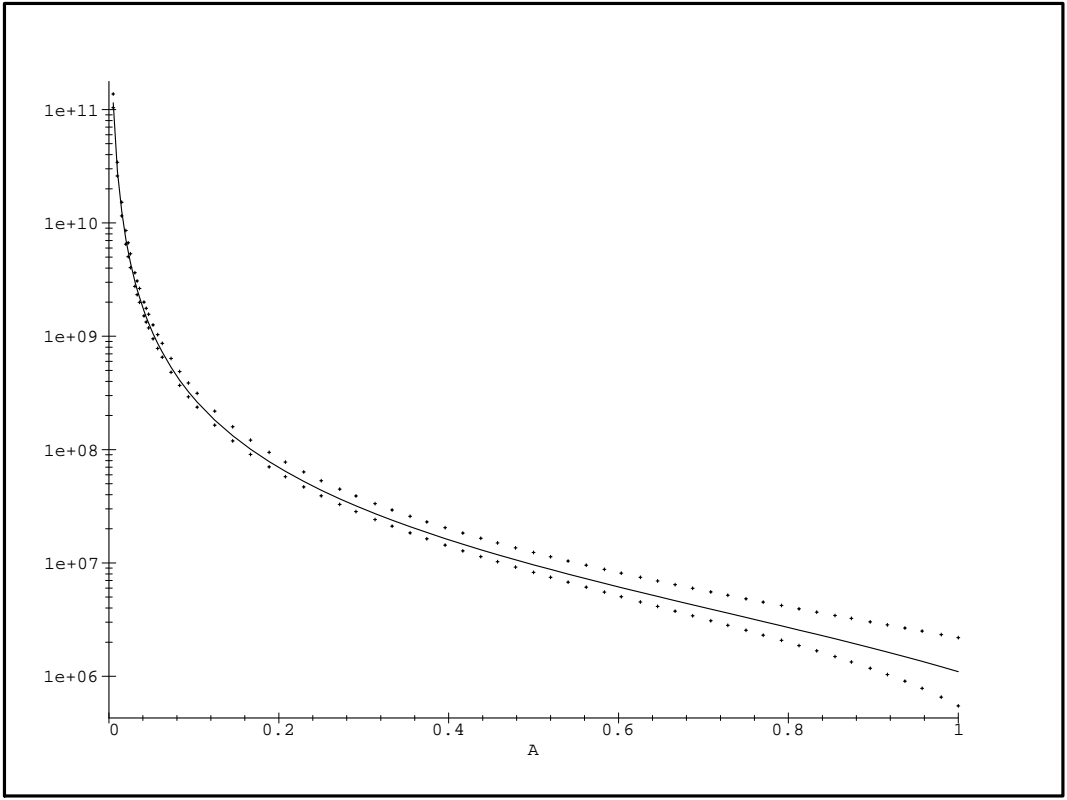}
\end{turn}
\caption{Four sigma discovery limit for $A_{cp}$ (X-axis)
as a function of luminosity 
for $B/S = 0.12,\ 0.24,\ 0.48$. The Y-axis is integrated luminosity in units
of nb$^{-1}$.}
\label{fig:sigacp}
\end{figure}

Here $\epsilon $ refers to the efficiency times $\Upsilon (4S)$
cross section, and
$\cal L$ to the integrated luminosity. 
We can now use Equation~\ref{eq:sigacpsqr} to compute the $4\sigma$ discovery
limit for $A_{cp}$ as a function of integrated luminosity.
This is shown in Figure~\ref{fig:sigacp}. 
We have assumed here
$\epsilon \times {\cal B}(B\to K^\pm\pi^\mp) = 0.7\times 10^{-5}$nb,
and $\rho = -0.008$.
The solid line is for $B/S = 0.24$. The dotted lines represent 
$B/S = 0.12$ and $B/S = 0.48$ respectively. Changing $\rho$ within a factor
of two makes no difference and is therefore not shown.

From Figure~\ref{fig:sigacp} we see that
with $\sim 300fb^{-1}$ we could establish $A_{cp}>0.0$ at the 
$4\sigma$ level if we measured $A_{cp} = 0.1$. The corresponding number for
$30$fb$^{-1}$ is $A_{cp} = 0.3$.

\begin{figure}
\centering
\leavevmode
\epsfysize=7cm
\epsfbox[45 150 525 630]{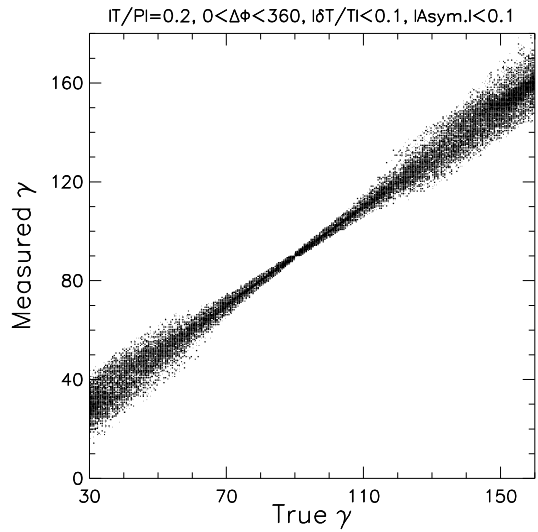}
\caption{Measured versus true $\gamma$ for $|A_{cp}|< 0.1$, $T_K/P_S=0.2$,
and $|\delta T_K/T_K|<0.1$.}
\label{fig:acpgamma}
\end{figure}

Figure~\ref{fig:acpgamma} shows the remaining
theoretical uncertainty on measuring $\gamma$ as discussed in the previous 
section if $|A_{cp}|<0.1$.

\section{Summary and Conclusion}
The design luminosities of PEP-II and CESR phase 4 are 
$30$fb$^{-1}$ and $300$fb$^{-1}$ per $10^7$ seconds, with PEP-II
starting operations in 1999, and CESR phase 4 being planned for 2003. 

We have evaluated experimental as well as theoretical errors for 
a determination of
Arg$(V_{ub})=\gamma $ at these future colliders, using 
branching fraction measurements for $B\to K^0_s\pi^\pm$,
$B\to K^+\pi^-$, $B\to K^-\pi^+$, $B\to \pi^\pm\pi^0$, and 
$d\Gamma/dq^2|_{q^2=m_\pi^2}$ for $B\to\pi l\nu$. The angle $\gamma$
is obtained from these measurements up to a fourfold ambiguity via 
construction of two amplitude triangles.

We project that $30$fb$^{-1}$
($300$fb$^{-1}$) of data taken at the $\Upsilon (4S)$ resonance should
result in an experimental error on $\gamma $ 
of $17^\circ$ ($8^\circ$). This estimate does not rely on any future
improvements to the existing CLEO analysis other than the 
$4\sigma$ particle ID separation between charged kaons and pions
that is expected for CLEO III as well as BaBar.

We find that the theoretical error on $\gamma $ depends strongly
on the geometry of the two amplitude triangles. 
For the experimentally preferred range of $\gamma$ ($41^\circ - 134^\circ$)
the theoretical errors tend to be small for small strong phase differences.
In particular, 
for strong phase differences
$|\Delta\phi|<20^\circ$ 
theoretical uncertainties of about $\pm 5^\circ$
for $ 70^\circ \le\gamma\le 130^\circ$ can be expected.
However, if we allow for large strong phase shifts theoretical errors
$2-3$ times as large are easily possible. 

Given the experimentally preferred range for $\gamma $, sizeable strong
phase shifts would result in direct CP violation large enough to be
easily measurable
with one year of ``nominal luminosity'' for
CESR phase 4 ($300$fb$^{-1}$). 

We find that $300$fb$^{-1}$ is sufficient to either measure
direct CP violation in $B\to K^\pm\pi^\mp$ or $\gamma$ to within
theoretical uncertainties comparable to the experimental errors. 

\vspace{0.7cm}
\centerline{\bf ACKNOWLEDGEMENTS}

We thank J.P. Alexander, L.K. Gibbons, M. Gronau, E. Thorndike,
as well as our collegues of the CESR phase 4 physics working group 
for useful discussions.
J.P. Alexander deserves special credit for many excellent comments on the
draft version of this paper.

This work was supported in part by the United States Department of Energy
under Contract No. DE-FG03-92-ER40701 and a Caltech Millikan Fellowship,
as well as the National Science Foundation.

\end{document}